\documentclass[pra,showpacs,twocolumn]{revtex4-1}
\newcommand{\vc}[1]{\mathbf{#1}}

\usepackage{psfrag}
\usepackage{bm,graphicx,amsmath}
\usepackage{color}
\begin{document}

\title{Quantum stochastic description of collisions in a 
canonical Bose gas}

\author{Patrick Navez$^1$ and Achilleas Lazarides$^2$}

\affiliation
{
$^1$Institut f\"ur Theoretische Physik, TU Dresden, 01062 Dresden, Germany\\
$^2$Max Planck Institute for the Physics of Complex Systems, 01187 Dresden, Germany
}

\date{\today}

\begin{abstract}
We derive a stochastic process that describes the kinetics of a one-dimensional Bose gas 
in a regime where three body collisions are important. 
In this situation the system becomes non integrable offering
the possibility to investigate  
dissipative phenomena  more simply compared to higher dimensional gases. 
Unlike the quantum Boltzmann equation 
describing the average momentum distribution, the stochastic approach allows a 
description of higher-order correlation functions in a canonical ensemble. 
As will be shown, this ensemble differs drastically from the grand canonical one. 
We illustrate the use of this method by determining the time evolution of the momentum 
mode particle number distribution and the static structure factor during the 
 evaporative cooling process.  
\end{abstract}


\maketitle

The stochastic kinetic description of a quantum gas has been 
the subject of considerable activity \cite{Steel,Proukakis,Strunz}. 
The main idea is to describe the state in terms of a set of stochastic 
classical fields $\psi^{(i)}(r,t)$ ($i=1,...,I$) associated to the bosonic 
quantum field $\psi(r,t)$ whose averages give access to the n-point 
correlation functions, possibly in 
the CN (canonical) ensemble. 
The determination of these quantities has been shown to be of relevance in 
recent interference experiments with cold atoms\cite{Schmiedmayer2} 
where mode probability functions have triggered considerable interest \cite{Zwerger,Demler}.  
As discussed in \cite{Steel}, the main 
difficulty is to be able to derive an associated Fokker-Planck(FP)-like equation 
for the probability in the P representation 
which can be recasted into a stochastic Langevin equation. Such a task 
cannot be achieved exactly since the resulting equation would be a non dissipative 
FP equation  \cite{Steel,Stoof,Drummond,Gardiner}. Only the use of 
the positive P representation appears to be tractable \cite{Drummond}.
Some approximations 
have to be  made in a bid to obtain numerically solvable dissipative equations. 
For the general case of finite temperature, only a phenomenological approach is available 
in which the stochastic terms 
guarantee the approach towards thermodynamic 
equilibrium without the requirement of an additional cutoff at large momentum 
\cite{Strunz}.

In this Letter, we present a microscopic derivation of a Langevin equation starting 
from the full many body Hamiltonian in which the stochasticity originates from 
collision processes. For simplicity, we study a one dimensional 
Bose gas subject to three-body interactions only (two-body interactions does not lead 
to effective thermalization in 1D). Such a system could be realized in cold atom experiments 
\cite{Schmiedmayer} and differs 
from the usual 1D exactly solvable Hamiltonian by the addition to the two-body quartic term 
of a sixtic interaction term resulting from virtual interaction with transverse degrees of freedom.
Its study allows simpler access to the exploration of
quantum collisions compared to the more involved higher dimensional 
systems \cite{ZNG,Navez,Proukakis}, 
in particular to the study of their enhancement due to the Bose statistics.
As an illustration, we consider the dynamics of evaporative cooling in the
CN ensemble and show that the time evolution of quantities such as correlations 
or mode probability distribution differ strongly from the one obtained 
from the Boltzmann equation valid 
only in the grand canonical (GN) ensemble.

Defining the atom mass $m$, the scattering length $\alpha_s$,
its 1D density $n$ and the transverse trap frequency $\omega_r$,  
we focus on the thermalization regime described in \cite{Schmiedmayer} 
characterized by a high transverse confinement 
$\hbar \omega_r \gg k_B T$, a weak correlation parameter $\gamma 
=2\alpha_s m \omega_r /\hbar n \ll 1$ in order to violate integrability 
so that the three-body collision rate overcomes the two-body collision rate responsible 
for transverse mode excitations.


Restricting ourselves to a uniform gas and defining the spatial Fourier components 
of the field   $\psi^{(i)}(r,t)=\sum_k \exp(ikr) \alpha^{(i)}_k (t)/\sqrt{L}$, we shall show how to derive 
the following Langevin-type equation:
\begin{eqnarray}\label{stoch}
d\alpha^{(i)}_p=(\Gamma_p^{in}-\Gamma_p^{out}+i\omega'_p)
\alpha_p^{(i)} dt 
+\sqrt{2 \Gamma_p^{in}}
d\eta_p^{(i)}
\end{eqnarray}   
where $\omega'_{p}$ is the kinetic and mean field energy of the atom,  
$\Gamma_p^{in}(t)$ and $\Gamma_p^{out}(t)$ correspond to the ingoing and outgoing collision 
terms respectively. These are averaged functionals of the $\alpha^{(i)}_{p}(t)$ and are determined 
from Eq.(\ref{in1},\ref{out1}).  
The noise $d\eta^{(i)}_p$ follows a Gaussian distribution with the only non trivial average 
$\langle d\eta_{p} d\eta_{p'}^*\rangle=\delta_{p,p'}dt$. In contrast  
to \cite{Proukakis,Stoof}, this equation includes both high and low 
energy modes in the stochastic process. 
The average is defined as
$\langle A \rangle= \sum_{i=1}^I  A^{(i)}/I$ for any set of realization $A^{(i)}$. It is done 
in the grand canonical (GC) ensemble and corresponds to the Monte-Carlo approximation of the 
integral over coherent state labeled by the $\alpha_p$'s \cite{Strunz}. In other words, 
the expectation value of 
any observable functional of the creation annihilation operator $a^\dagger_p$ and $a_p$, 
is in the limit of a large set $I$, 
identical to an average over 
the stochastic variables $\alpha^*_p$ and $\alpha_p$. 
 
We note the connection with the quantum Boltzmann equation. Multiplying (\ref{stoch}) 
by its complex conjugate and taking the average over the ensemble, the stochastic equation is 
connected to the quantum Boltzmann equation
for the momentum distribution $\overline{n}_k= \langle n_k \rangle 
=\langle |\alpha_k|^2 \rangle$:
\begin{eqnarray}\label{boltzmann}
\frac{d \overline{n}_k(t)}{dt}= 2{\Gamma}^{in}_k (1+\overline{n}_k)-
2{\Gamma}^{out}_k\overline{n}_k 
\end{eqnarray}
This equation becomes closed if 
the  ${\Gamma}^{in}_k$ and  ${\Gamma}^{out}_k$ are functionals of $\overline{n}_k$.
This is realized through the application of the Wick's decomposition 
theorem (which corresponds to the stosszahlansatz): 
defining the product $M_n(\{\alpha_k\}) =\alpha_{k_1}^* \dots \alpha_{k_n}^* 
\alpha_{k'_1} \dots \alpha_{k'_{n}}$, its average in the GC ensemble for a uniform gas 
is decomposed as:
\begin{eqnarray}\label{3}
\langle M_n(\{\alpha_k\})\rangle = \overline{n}_{k_1}\dots \overline{n}_{k_n}
\sum_{\{k_j\} \in P} \prod_i \delta_{k_i,k_j}
\end{eqnarray}
where $P$ is the permutation ensemble. 
On the contrary, the stosszahlansatz is not necessary anymore in Eq.(\ref{stoch}) and 
correlations can be taken into account. 

Another advantage of the stochastic formalism 
is the possibility to relate the averages in the CN ensemble to the GC ones by means of 
the weight functions $W_N(\{\alpha_k\})=e^{-\sum_k |\alpha_k|^2}\left(\sum_k |\alpha_k|^2\right)^N/N!$ 
so that for $N$ atoms we obtain for any product \cite{Strunz}:
\begin{eqnarray}\label{Mn}
\langle M_n(\{\alpha_k\}) \rangle_{N} = 
\frac{\langle 
W_{N-n}(\{\alpha_k\})M_n(\{\alpha_k\})\rangle}
{\langle W_{N}(\{\alpha_k\}) \rangle}
\end{eqnarray}
The weight function plays the role of a projection operator restricting 
the particle number to $N$.
Probability distribution in the CN ensemble can also be determined for mode population. 
For example, defining the weight functions $W_{N,k\not=0}=
e^{-\sum_{k\not=0} |\alpha_k|^2}\left(\sum_{k\not=0} |\alpha_k|^2\right)^N/N!$ 
and $W_{N,k=0}=
e^{-|\alpha_0|^2}|\alpha_0|^{2N}/N!$, 
the probability distribution for the zero momentum mode reads:
\begin{eqnarray}\label{Pn0}
P_N(n_0)=  \frac{\langle 
W_{N-n_0,k\not=0}(\{\alpha_k\})W_{n_0,k=0}(\{\alpha_k\})\rangle}
{\langle W_{N}(\{\alpha_k\}) \rangle}
\end{eqnarray}
\begin{figure}[t]
\includegraphics[width=8cm]{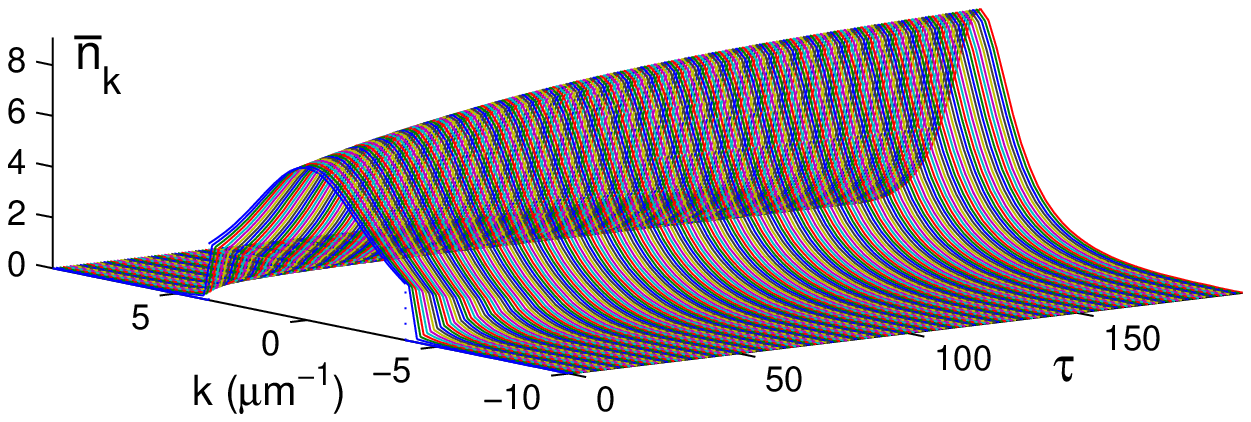}
\includegraphics[width=8cm]{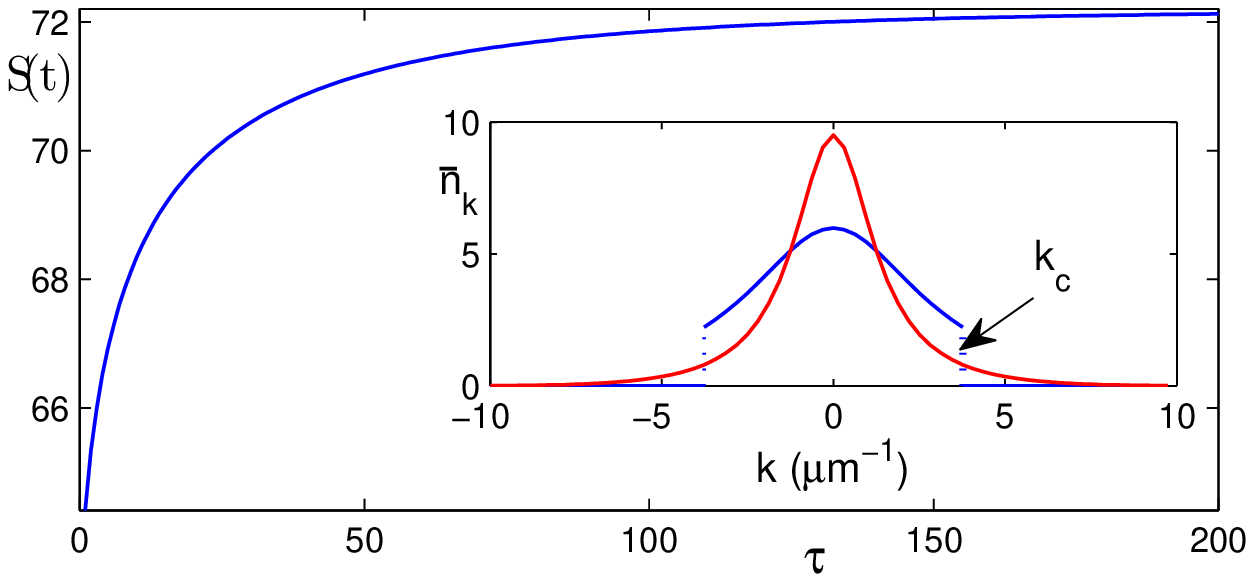}
\caption{Evolution of the momentum average distribution $nk$ vs. the reduced time $\tau$. 
The corresponding entropy is shown below and indicate how the distribution is 
close to equilibrium. Initial and final distributions are shown in blue and red respectively 
in the inset. }
\label{fig1}
\end{figure}
The determination of the stochastic equation is done as follows. 
We start from the ternary many body Hamiltonian \cite{Schmiedmayer}:
\begin{eqnarray}
H=\sum_{{k}} \omega_{k} 
c^\dagger_{k} c_{k} 
-\frac{g_{3}}{L^2}\sum_{\sum_i p_i =\sum_i q_i}
\!\!\!\!\!\!\!
c^\dagger_{p_1} c^\dagger_{p_2}
c^\dagger_{p_3}
c_{q_1} c_{q_2}
c_{q_3}
\end{eqnarray}
where $\omega_k=k^2/2m$ is the atom kinetic energy and
$g_{3}=2\ln(4/3) \hbar \omega_r \alpha_s^2 $ is 
the ternary three body interactions. The suppression of the 
quartic interactions proportional to 
$g_2= \hbar \omega_r \alpha_s$ is valid in a region 
where the mean field interaction energy is negligible compared to 
the kinetic energy i.e. $g_2 n \ll k_B T$ so that the phonon excitation 
energy are mostly particle-like \cite{Schmiedmayer}. 
Starting the density matrix $\rho(t)$, we obtain the reduced 
density matrix for the momentum mode ${p}$ by taking the trace 
over all other modes:
\begin{eqnarray}
\rho_{{p}}(t)={\rm Tr}_{\setminus {p}}
\left(\rho(t)\right)
\end{eqnarray}
The stochastic time evolution of the mode
${p}$ is derived by considering the other modes as a bath for this mode. 
For this purpose, we decompose the density matrix in terms of an uncorrelated  
contribution and a correlated one:
\begin{eqnarray}
\rho(t)= \prod_{{p}} \rho_{{p}}(t)
+\delta \rho(t)
\end{eqnarray}
The Hamiltonian can be decomposed into three terms:
\begin{eqnarray}
H=H_1 + H_2 +H_3 
\end{eqnarray}
where $n_k=a^\dagger_k a_k$ and
\begin{eqnarray}
H_1=  \omega_{p} n_p + {\cal O}(1/L)
\end{eqnarray}
\begin{eqnarray}
H_2= A_{{p}} n_p
+C_{p}^\dagger a_{p} + 
a^\dagger_{{p}}C_{p}+ {\cal O}(1/L)
\end{eqnarray}
and where $A_{{p}}$, $C_{p}$ and $H_3$ are operators describing the other modes. We 
omit terms of the order ${\cal O}(1/L) \ll n$ which are negligible 
in the thermodynamic limit. 
Using the formalism in \cite{Gardiner} to derive the master equation up to the second order 
in $g_3$, we obtain:
\begin{eqnarray}\label{masterg} 
\frac{\partial \rho_{p}}{\partial t}
&=& -\frac{i}{\hbar} [H'_1,\rho_{p}]+
\Gamma_{p}^{in}(t)
(2a^\dagger_{p}\rho_{p}a_{p}-
[a_{p}a^\dagger_{p}, \rho_{p}]_+)
\nonumber \\ 
&-&\Gamma_{p}^{out}(t)
([a^\dagger_{p}a_{p}, \rho_{p}]_+
-2a_{p}\rho_{p} a^\dagger_{p})
\end{eqnarray}
where 
\begin{eqnarray}
H'_1=H_1+
{\rm Tr}_{\setminus {p}}(\prod_{{k}\not= {p}}
\vc{\rho}_{{k}}
H_2)
=\omega'_p n_p
\end{eqnarray}
defining $\omega'_p=\omega_p- 3!g_3 n^2$ and $n=\sum_k \langle n_k \rangle/L$ 
and where
\begin{eqnarray}
\Gamma_{p}^{in}(t)
&=&{\rm Re}\frac{1}{\hbar^2}
\int_0^\infty dt'
\langle C^\dagger_{{p}}(t')C_{p}\rangle
\nonumber \\ \label{in}
&=&{\rm Re}\frac{1}{\hbar^2}
\int_0^\infty dt'
{\rm Tr}(C^\dagger_{{p}}(t')C_{p} 
\prod_{{k}\not= {p}}\rho_{k}(t)) 
\\ \label{out}
\Gamma_{p}^{out}(t)
&=&{\rm Re}\frac{1}{\hbar^2}
\int_0^\infty dt'
\langle C_{{p}}(t')C^\dagger_{p} \rangle
\end{eqnarray}
Terms containing the operator $A_p$ do not contribute up 
to $g^2_3$ by symmetry between the in and the out terms.  
The only contribution comes from the operator $C_{p}$ which expressed in the interaction 
picture becomes: 
\begin{eqnarray}
C_{p_1}(t)= \frac{3g_{3}}{L^2}\!\!\!\!\!\!\!\!\!\sum_{\hspace{0.5cm}\sum_i p_i =\sum_i q_i}
\!\!\!\!\!\!\!\!\!\! e^{i\sum_i(\omega_{p_i} - \omega_{q_i})t} c^\dagger_{p_2}
c^\dagger_{p_3} c_{q_1} c_{q_2} c_{q_3}
\end{eqnarray}
The stochastic equation can be implemented numerically directly using Eq.(\ref{in},\ref{out}) 
but in order to avoid the summation over too many momentum variables we make the random phase 
approximation. For a uniform gas, we can neglect off-diagonal contributions in the $p_i$ and  
$q_i$  since they induce a phase factor that appears to be random and thus cancels 
in the summation process. Thus, only 
diagonal components remain. Carrying out the integral over $t'$, we obtain with the 
following expression: 
\begin{widetext}
\begin{eqnarray}\label{in1}
\Gamma_{p}^{in}(t)
&=&(3!)^2\pi \left(\frac{g_3}{\hbar L^2}\right)^2
\sum_{\{{p_i},
{q_i}\}}
\delta_{\sum_i {p_i},
\sum_i {q_i}}
\delta(\sum_i \omega_{p_i}-\sum_i \omega_{q_i})
\langle n_{p_2}n_{p_3}
(n_{q_1}+1)(n_{q_2}+1)(n_{q_3}+1)\rangle
\\ \label{out1}
\Gamma_{p}^{out}(t)
&=&(3!)^2 \pi \left(\frac{g_3}{\hbar L^2}\right)^2
\sum_{\{{p_i},
{q_i}\}}
\delta_{\sum_i {p_i},
\sum_i {q_i}}
\delta(\sum_i \omega_{p_i}-\sum_i \omega_{q_i})
\langle n_{q_1}n_{q_2}n_{q_3}
(n_{p_2}+1)(n_{p_3}+1)\rangle
\end{eqnarray}
Note the Bose enhancement factor (terms $n_{p}+1$) that amplifies 
the collision process when the output modes are already populated.   
From this form we can deduce the FP equation associated to  
the master equation (\ref{masterg}):
\begin{eqnarray}\label{Pq}
\frac{\partial P_{p}(t)}{\partial t}=\biggl \{
-i\omega'_{p}
\left(\alpha_{p}\frac{\partial}{\partial \alpha_{p}}
-\alpha^*_{p}\frac{\partial}{\partial \alpha^*_{p}}\right)
+ (\Gamma_{p}^{out}(t)-\Gamma_{p}^{in}(t))
\left(\frac{\partial}{\partial \alpha_{p}}\alpha_{p}
+\frac{\partial}{\partial \alpha^*_{p}}\alpha^*_{p}\right)
+2\Gamma_{p}^{in}(t)\frac{\partial}{\partial \alpha_{p}}
\frac{\partial}{\partial \alpha^*_{p}}\biggr \} P_{p}(t)
\end{eqnarray}
\end{widetext} 
from which we deduce the Langevin equation Eq.(\ref{stoch}).
The solution of this equation is a Gaussian distribution
$P_p (t)= \exp(-|\alpha_p|^2/ \overline{n}_p(t))/\overline{n}_p(t)$ 
with the normalization 
$\int d^2 \alpha_p  P_p (t)/\pi=1$ and where $\overline{n}_p(t)$ has 
to fulfill Eq.(\ref{boltzmann}). 
In equilibrium in the GC ensemble, through the application of 
Eq.(\ref{3}) on Eq.(\ref{in1},\ref{out1}), we recover the Bose-Einstein 
distribution $\overline{n}_k=1/[\exp((\omega_{{p}}-\mu)/k_B T)-1]$ 
as the stationary solution where the parameter $\mu$ defines  
the chemical potential. 

For an illustration of all these considerations, we apply 
the stochastic method to the process of evaporative cooling. We take  
a gas of $^{87}Rb$ of $n= 5.25\,\mu m^{-1}$ confined in a box of size $L=20\mu m$ 
with $\omega_r/2\pi= 6kHz$ and $\alpha_s=5.3nm$. We choose the dimensionless time 
$\tau =10^6 t/t^*$ where $1/t^*= (3!)^3\left(g_3/\hbar L\right)^2 m/\pi \hbar$.
  
We start 
from an initial Bose gas at high temperature $T_i=29nK$, we 
then remove the hottest atom with 
momentum higher than $k_c=4\mu m^{-1}$ and study the relaxation process of this cut distribution 
towards an equilibrium one with a lower temperature $T_f=9nK$ 
at a time estimated to $t=0.14 s$ ($\tau=200$). The time 
evolution of the GC distribution calculated from 
Boltzmann approach of Eq.(\ref{boltzmann}) is shown in Fig.1 together 
with the entropy $S(t)=\sum_p (\overline{n}_p+1)\log(\overline{n}_p+1)- \overline{n}_p 
\log(\overline{n}_p)$ evolution that allows to monitor the speed at which the equilibrium 
state is reached. Its production can be shown to be always positive and stops 
at equilibrium \cite{Balescu}.  

\begin{figure}[t]
\includegraphics[width=9cm]{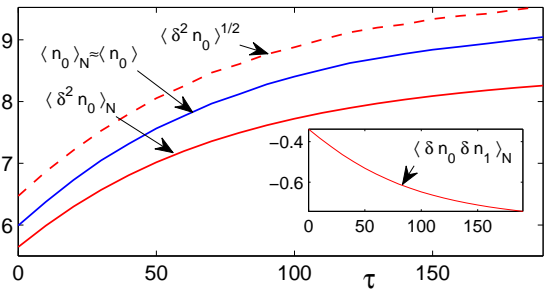}
\caption{Evolution of the CN and GC  mean atom number in the zero mode and their 
fluctuations vs. the reduced time $\tau$. The negative correlation with the first 
excited mode is shown in the inset.}  
\label{fig3}
\end{figure}
\begin{figure}[t]
\includegraphics[width=8cm]{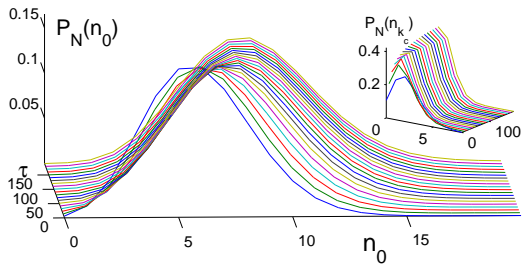}
\caption{Evolution of the CN atom probability distribution function in the zero mode 
vs. the reduced time $\tau$. For comparison, the distribution for the 
mode $k_c$  is represented in the inset.}  
\label{fig4}
\end{figure}

The stochastic approach allows a more refined description in the GC ensemble. 
For instance, the atom probability distribution for the momentum mode 
is determined 
from the solution of Eq.(\ref{Pq}) and corresponds to a Poisson distribution:
\begin{eqnarray}
P(n_p)=\!\!\int  \!\!\frac{d^2 \alpha_p}{\pi}\frac{|\alpha_p|^{2n_p}}{n_p!}e^{-|\alpha_p|^2}\!
P_q=\frac{\overline{n}_p^{n_p}}{(\overline{n}_p+1)^{n_p+1}} 
\end{eqnarray}
with the larger fluctuations $\langle \delta^2 n_0 \rangle_{GC}=\overline{n}_0(\overline{n}_0+1)$.
Correlations between modes are non existent e.g. 
$\langle \delta n_0 \delta n_1\rangle _{GC}=0$ where $n_1$ is the mode 
for the first excited state $k=2\pi/L$.

The situation changes drastically in a canonical description as the fixed atom number 
restricts the possibility of fluctuations. Using the formulae 
(\ref{Mn}) and (\ref{Pn0}) over a sample of $10^6$ variables for the 
stochastic process (\ref{stoch}), if the average value remains practically unchanged, 
the fluctuations are significantly reduced as shown as in Fig.2.  The process of evaporative 
cooling has the effect to increase these fluctuations together with the mean population.
The negative values of $\langle \delta n_0 \delta n_1\rangle_{CN}$
is an evidence of correlations between modes in the CN ensemble. 

\begin{figure}[t]
\includegraphics[width=8cm]{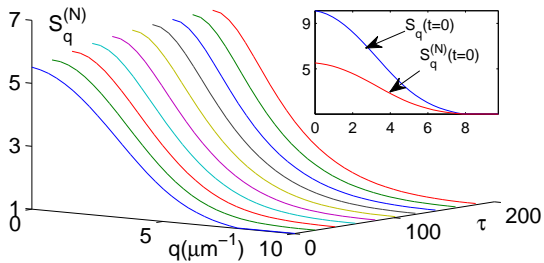}
\caption{Evolution of the static structure factor $S_q^{(N)}$ 
in the CN ensemble. For comparison, the corresponding GC function $S_q(t=0)$ is represented in the inset.}  
\label{fig5}
\end{figure}

The canonical probability distribution resembles a Gaussian one. Fig.3
shows how this distribution evolves with time and 
illustrates the benefit of using a stochastic approach for 
quantities that differ according ensembles. 
For the mode $k_c$, the distribution is initially peaked and becomes monotonic during 
the time evolution. 

The static structure factor is defined in GC and CN ensembles respectively as \cite{Stringari}:
\begin{eqnarray}
S_q= \langle \delta \rho_q \delta \rho_{-q} \rangle/N \quad 
S_q^{(N)}= \langle \delta \rho_q \delta \rho_{-q} \rangle_N/N \quad 
\end{eqnarray}
and corresponds to the density fluctuations $\rho_q=\sum_k c^\dagger_{k-q} c_{k}$.
These functions can be measured in Bragg spectroscopy experiments 
\cite{Ketterle} and  reach unity for large momentum $q$.
In Fig.\ref{fig5}, we determine the time evolution of these fluctuations during 
the evaporative cooling process. We notice that the presence of correlation in the CN ensemble 
reduces significantly the static structure factor.     


In conclusion, we have reformulated kinetic theory in terms of a stochastic approach where 
the stochastic variables describe the various momentum modes of the gas. 
Through a specific weight average, the CN description becomes accessible and shows
drastic differences in comparison with the GC one.
The method has been used for a one-dimensional gas but is quite 
general and can be applied to more general cases of higher dimension, including 
higher order interaction terms or in presence of longitudinal trap confinement.

PN would like to thank 
S. Heller, H. Hauptmann, S. Kr\"onke and W. Strunz   
for  helpful discussions.


\begin{thebibliography}{9}

\bibitem{Steel}
M. J. Steel, M. K. Olsen, L. I. Plimak, P. D. Drummond, S. M. Tan, M. J. Collett, D. F. Walls, and R. Graham
Phys. Rev. A {\bf 58}, 4824–4835 (1998).
\bibitem{Proukakis}
Proukakis N.P. and Jackson B., J. Phys. B: At. Mol. Opt. Phys. {\bf 41} 
203002 (2008); S. P. Cockburn1, A. Negretti, N. P. Proukakis, and C. Henkel, 
Phys. Rev. A {\bf 83}, 043619 (2011).
\bibitem{Strunz}
S. Heller and W.T. Strunz, 
J. Phys. B {\bf 42}, 081001 (2009); S. Heller's thesis.
\bibitem{Schmiedmayer2}
S. Hofferberth et al., Nature Physics {\bf 4}, 489 (2008)
\bibitem{Demler}
V. Gritsev, E. Altman, E. Demler, A. Polkovnikov,
Nature Physics {\bf 2}, 705 (October 2006).
\bibitem{Zwerger}
S.F Rath and W. Zwerger, Phys. Rev. A {\bf 82}, 053622 (2010).
\bibitem{Stoof} H.T.C. Stoof, Phys. Rev. Lett. {\bf 78}, 768 
(1997). 
\bibitem{Gardiner}
C. Gardiner, {\it Quantum Noise} (Springer  Verlag, 1990).
\bibitem{Drummond}
P. Deuar and P.D. Drummond, Phys. Rev. Lett. {\bf 98}, 120402 (2007);
P. Deuar and P.D. Drummond, Phys. Rev. A {\bf 66}, 033812 (2002);
P.D. Drummond, Phys. Rev. Lett. {\bf 92}, 040405 (2004). 
\bibitem{Schmiedmayer}
I. E. Mazets, J. Schmiedmayer, 
New J. Phys. {\bf 12}, 055023 (2010).
\bibitem{ZNG}
A. Griffin, T. Nikuni, E. Zaremba, 
{\it Bose-condensed gases at finite temperatures} 
(Cambridge University Press, 2009). 
\bibitem{Navez} Patrick Navez, 
Physica A {\bf 356}, 241-278 (2005). 
\bibitem{Balescu}
R. Balescu,
{\it Equilibrium and nonequilibrium statistical mechanics}
(New York, Wiley-Interscience, 1975).
\bibitem{Stringari}
L.  Pitaevski, S. Stringari, {\it Bose-Einstein condensation} 
(Clarendon Press, 2003).
\bibitem{Ketterle}
Stamper-Kurn et al., Phys. Rev. Lett. {\bf 83}, 2876 (1999).
\end{thebibliography}
\end{document}